\documentclass[11pt,a4paper]{article}

\usepackage[T1]{fontenc}
\usepackage{newtxtext}
\usepackage{newtxmath}
\usepackage[
    top=1in,
    bottom=1in,
    left=1in,
    right=1in
]{geometry}
\usepackage{amsmath}
\usepackage{array}
\usepackage{booktabs}
\usepackage{caption}
\usepackage{graphicx}
\usepackage{hyperref}

\hypersetup{
    colorlinks=true,
    linkcolor=blue,
    citecolor=blue,
    urlcolor=blue
}

\newcounter{constraintterm}
\newcounter{objectiveterm}
\newcommand{\Ctag}{\refstepcounter{constraintterm}\tag{C.\arabic{constraintterm}}}
\newcommand{\Otag}{\refstepcounter{objectiveterm}\tag{O.\arabic{objectiveterm}}}

\title{Quantum Resource Estimation for Minimising Energy Grid Losses}
\author{%
Camille de Valk$^{1,2,3}$, Milou van Nederveen$^1$, Koen Reerink$^2$, Werner van Westering$^1$\\[0.5em]
\normalsize $^1$Alliander, Arnhem, The Netherlands\\
\normalsize $^2$Capgemini's Quantum Lab, Utrecht, The Netherlands\\
\normalsize $^3$Applied Quantum Algorithms, Leiden University, Leiden, The Netherlands\\[0.25em]
\normalsize \texttt{camille.de.valk@alliander.com}
}
\date{}

\begin{document}

\maketitle

\begin{center}
\small
This paper is a preprint of a paper accepted by the Proceedings of the CIRED 2026 Brussels Workshop and is subject to Institution of Engineering and Technology Copyright. When the final version is published, the copy of record will be available at IET Digital Library.
\end{center}

\begin{center}
\textbf{Keywords:} QUANTUM COMPUTING, DISTRIBUTION NETWORK RECONFIGURATION, POWER LOSS REDUCTION, HIGHER-ORDER UNCONSTRAINED BINARY OPTIMISATION (HUBO)
\end{center}

\begin{abstract}
Distribution network reconfiguration (DNR) can minimise power losses by identifying the optimal topology of the electricity grid. Determining the minimum loss configuration is NP-hard, and classical optimisation methods struggle to scale to real-world distribution grids. This paper explores the use of gate-based quantum computing to solve DNR for power loss reduction. We formulate DNR as a higher-order unconstrained binary optimisation (HUBO) problem, avoiding the need for auxiliary variables, thereby reducing the required number of qubits. This is applied to a real medium voltage (MV) network operated by Alliander, a Dutch distribution system operator (DSO). For each biconnected component in the network graph, we construct the corresponding HUBO, derive the cost and mixer operators, and determine the number of required logical qubits and rotation gates. These are then mapped to physical qubits and execution time estimates using quantum resource estimation (QRE). The results suggest that the quantum resource requirements depend not only on component size but also on structural characteristics such as connectivity and cyclicity. Overall, the novelty of this work lies in directly framing the optimisation problem as a HUBO, applying it to real-world MV networks, and performing a QRE to assess future feasibility.
\end{abstract}

\section{Introduction}
Reconfigurable electricity grids contain switches that can be opened and closed, changing the topology of the network. Although this capability was traditionally intended for fault detection and isolation, distribution network reconfiguration (DNR) also enables power loss reduction \cite{lotfi2024power}. Reducing losses can help relieve grid congestion, lower carbon emissions, and reduce operational costs. The impact of loss-minimising DNR is significant for distribution system operators (DSOs), many of which operate increasingly loaded networks with a growing amount of distributed renewable energy. In the Netherlands, Alliander DSO supplies the electricity to around six million customers. Alliander's medium-voltage (MV) network has a meshed infrastructure but is operated radially using switchable links, making it an ideal candidate for reconfiguration. A previous study has indicated that DNR can reduce losses in Alliander's MV by 15-27\% \cite{werner-cired}. Based on internal analysis, such improvements could save millions of euros and reduce carbon emissions.

\begin{figure}[t]
\centering
\includegraphics[width=0.6\textwidth]{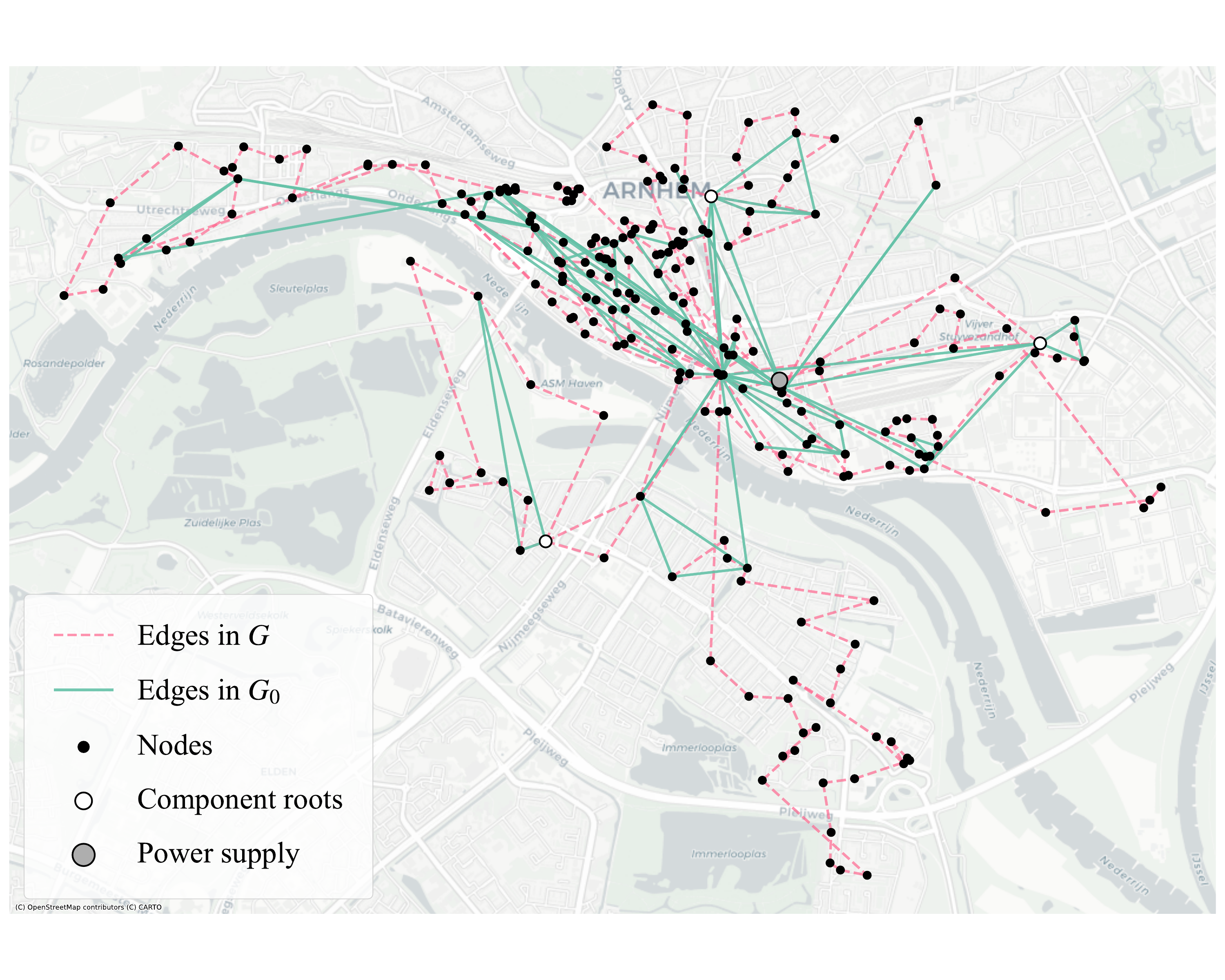}
\caption{The MV network in Arnhem, shown with its graph topology. The network is decomposed into biconnected components $G_C$ (pink dashed lines) and their topological minors $G_0$ (blue lines). Normal nodes appear as black dots, component roots of each $G_0$ as white dots, the power supply as a grey dot. Only nodes with known coordinates are plotted.}
\label{fig:arnhem-MV}
\end{figure}

Determining the optimal DNR configuration to minimise power losses proves to be computationally challenging. Mathematically, this problem has been proven to be NP-hard \cite{werner-cired}. Classical optimisation techniques, while effective for small or moderately sized networks, struggle with the combinatorial complexity of modern grid reconfiguration problems. Despite decades of research, there is no ideal solution method \cite{lotfi2024power}. Quantum computing has the potential to advance optimisation in multiple ways, such as improving solution quality and efficiency \cite{abbas2024challenges}. In this study, we explore the possibilities of using quantum computing to find the DNR for minimising power losses.

Recent work formulated DNR for a quantum annealer as a quadratic unconstrained binary optimisation (QUBO) problem for a 33-node IEEE test network \cite{silva_qubo_2023, silva_quantum_2023}. In that approach, (higher-order) constraints are implicitly transformed to QUBO form. Because we focus on gate-based quantum computing, we model all constraints explicitly as higher-order penalty terms. Hence, we use a more generic formulation, namely higher-order unconstrained binary optimisation (HUBO) without auxiliary variables. Directly solving the HUBO problem can reduce qubit count, bringing practical gate-based implementation closer \cite{romero_bias-field_2025}.

Additionally, we focus on real-world systems by evaluating the practical feasibility of optimising the topology of an MV network of Alliander, namely Arnhem, as shown in Figure \ref{fig:arnhem-MV}. The graph is decomposed into parts, which can each be solved independently. We create a HUBO formulation and using a cost and mixer operator, translate this to the number of logical number of qubits and rotation gates. To gain insights into the feasibility of utilising this in the near future, the resource requirements of the quantum algorithm are estimated. This is referred to as quantum resource estimation (QRE), and given the estimated runtimes and physical number of qubits, can be used to gain insights into the future of using quantum computers for the operational tasks of DSOs. To compare our work with \cite{silva_qubo_2023}, we also apply this approach to the IEEE-33 test network \cite{ieee-33}, and present the results.

\section{Methodology}
Following \cite{silva_qubo_2023}, we formulate network reconfiguration as an unconstrained binary optimisation problem. We model the distribution grid as a directed graph $G = (V, E)$, where vertices $n\in V$ represent substations, loads, or junctions with currents $I_L^n$, and edges $(u,v)\in E$ represent distribution lines with resistance $R_{u,v}$, load limits and voltage limits. Edges are assumed to be switchable, enabling network reconfiguration, while operation is constrained to a radial, tree-like configuration.

The goal of our optimisation problem is to find a network topology $T^*$ that minimises the sum of all load losses.
As operational constraints impose radiality of the network, any feasible configuration of the network is necessarily a spanning tree of $G$.
Formally, we solve
\begin{equation}
    T^* = \underset{T\in ST(G)}{\mathrm{arg}\text{ }\mathrm{min}} \sum_{(u, v)\in E(T)}L_{u,v}(T),
    \label{eq:original-objective}
\end{equation}
with $ST(G)$ the set of all spanning trees of $G$, $E(T)$ the edges in the spanning tree $T$, $I_{u,v}(T)$ the current flowing through edge $(u,v)$ in the spanning tree $T$, and $L_{u,v}(T) \equiv |I_{u,v}(T)|^2R_{u,v}$ the load on edge $(u,v)$ in the spanning tree $T$. The current $I_{u,v}(T)$ is given by the sum of the currents of all downstream nodes $D_{u,v}(T, v_0)$ as
\begin{equation}
    I_{u,v}(T) = \sum_{n\in D_{u,v}(T, v_0)}I_n^L,
\label{eq:I_uv(T)}
\end{equation}
because currents are additive.

The datasets used in this work, presented in Table \ref{tab:hubo_qaoa_biconnected}, include a benchmark network \cite{ieee-33} and a real-world distribution network from Alliander DSO. From these datasets, we extract the graph topologies. While electrical parameters such as resistance and voltage limits are essential for solving the optimisation problem, they are not required for resource estimation, which only needs the number of variables and interaction terms.

To reduce the problem size, we follow \cite{silva_qubo_2023} and decompose $G$ into $n$ biconnected components $\{G_{C_0}, G_{C_1}, \ldots, G_{C_n}\}$. We retain only non-trivial biconnected components, i.e., components containing at least one cycle. The biconnected components can be solved independently, because the configuration upstream of a component root $m$ cutting the biconnected components does not affect the load at node $m$, see equation \eqref{eq:I_uv(T)}. We disregard dyads and tree-like substructures, as they admit a unique feasible configuration and are therefore trivial to solve. Nodes with identical edge sequences are merged.

Let $G_C$ be such a non-trivial biconnected component. We reduce $G_C$ to its topological minor $G_0$ through edge lifting. More specifically, for degree-2 nodes, if there's no edge between their neighbours, we place a new edge between their neighbours and remove the degree-2 node. This preserves the cycle and path structure from $G_C$, yielding equivalent spanning trees (up to lifted edges), while reducing the number of nodes and edges.

\subsection{HUBO-formulation of this problem}
In contrast to previous work \cite{silva_qubo_2023}, we model the distribution network reconfiguration problem as a higher-order unconstrained binary optimisation (HUBO) problem. In general, a HUBO can be expressed as
\begin{equation}
H = \alpha
+ \sum_i \alpha_i x_i
+ \sum_{i,j} \alpha_{ij} x_i x_j +...
+ \sum_{i,\ldots,n} \alpha_{i,\ldots,n} x_i x_j \cdots x_n,
\end{equation}
where $x_i \in \{0,1\}$ are binary decision variables and the coefficients $\alpha$, $\alpha_i$, $\alpha_{ij}$, and $\alpha_{i,\ldots,n}$ represent cost contributions arising from interactions of increasing order. Binary variables encode both network topology decisions and auxiliary quantities used in the objective and constraints. The final cost function is a weighted sum of penalty terms that enforce feasibility and an objective term representing power losses in the distribution grid. Avoiding a reduction to a QUBO formulation preserves the higher-order structure of the problem and gives a more direct representation of the underlying physical and topological constraints.

\subsubsection{Modelling constraints as penalties in HUBO}

In our novel HUBO formulation, constraints are modelled explicitly as penalties, rather than in the QUBO formulation, where constraints are implicitly handled by D-Wave's SDK \cite{silva_qubo_2023}. We model three types of constraints: linear sum constraints, interaction constraints, and implies constraints.

\subsubsection{Linear sum constraints}
In a linear sum constraint, a linear sum of variables equals one, i.e., $\sum_i x_i=1$. This can be modelled as a penalty term
\begin{equation}
    c = \left( \sum_{i} x_{i} - 1 \right)^2.
\end{equation}
We employ this penalty to model vertex constraints.

\subsubsection{Interaction constraints}
An interaction constraint forbids the interaction between multiple variables, i.e., $\prod_{i, ..., n} x_i=0$. In our formulation, that interaction can be directly added to the HUBO as
\begin{equation}
    c_{i,...,n}=\prod_{i, ..., n} x_i.
\end{equation}
This type of penalty is used for edge and cycle constraints.

\subsubsection{Implies constraints}
In the formulation of the network reconfiguration problem, we encounter constraints of the form $x_i \implies x_j$ or $x_i\implies \neg x_j$, with $\neg x_j \equiv 1-x_j$, since we are dealing with only binary variables. These constraints can be added as a penalty by using
\begin{align}
    c_{x_i \implies x_j} &= x_i - x_i x_j \label{eq:implies-pos}\\
    c_{x_i \implies \neg x_j} &= x_i - x_i(1-x_j)=x_ix_j.  \label{eq:implies-neg}
\end{align}
If $x_i=1$, the penalty enforces $x_j=1$ in the positive case, \eqref{eq:implies-pos}, and $x_j=0$ in the negative case, \eqref{eq:implies-neg}.
If $x_i=0$, $x_j$ is unconstrained.
The implies constraint is used to link edge variables to virtual edge variables, to construct path variables, and to construct load-arc variables.

\subsubsection{Penalties in the HUBO}
With the novel method of modelling constraints for the network reconfiguration problem as penalties in the HUBO, we can combine all constraints to form the first part of the HUBO. We list all constraints and refer to \cite{silva_qubo_2023} for full definitions before reformulation as higher-order penalties.

\begin{itemize}
    \item \textit{\textbf{Vertex constraints}} Vertex constraints ensure valid local connectivity of the network. For every non-root vertex $v \in V \setminus \{v_0\}$, exactly one incident edge must be active.
    \item \textit{\textbf{Edge Constraints}} Edge constraints enforce consistency in the directed representation of the network. Each undirected edge $(u,v)$ is modelled by two directed arc variables $e_{u,v}$ and $e_{v,u}$, of which at most one may be active.
    \item \textit{\textbf{Cycle constraints}} To enforce global radiality, the active subgraph must be cycle-free. This requirement cannot be guaranteed by local degree or edge-direction constraints alone, as directed cycles may still occur in the selected subgraph. Cycle constraints are constructed using a \emph{cycle basis} of the reduced graph, starting from facial cycles and adding virtual edges to prevent the closure of combinations of cycles as necessary.
    \item \textit{\textbf{Path constraints}} Path constraints ensure that every vertex along a path $P_{u,v}$ in the network $G_C$ is consistently connected to the root node $v_0$, preventing disconnected configurations that would otherwise satisfy cycle constraints but violate radiality. This allows the optimisation to effectively find a spanning tree on $G_0$, while the full network $G_C$ is considered for the loads.
\end{itemize}

\subsubsection{Objective Function}
The objective of the HUBO formulation is to minimise total ohmic power losses under a constant-current load approximation, see equation \eqref{eq:original-objective}. The penalty terms constrain all variables $e_{u,v}$ to spanning trees $T\in ST(G)$, so minimising the sum of loads will yield the optimal tree $T^*$. Because a tree on $G_0$ is also a tree on $G_C$ (up to lifted edges), we only need to consider edges $E_0$, i.e., equation \eqref{eq:original-objective} can be written as $\sum_{(u,v)\in E_0} L_{u,v}^0$ and
\begin{equation}
    L_{u,v}^0=L_{u,v}^D+L_{v,u}^D,
\end{equation}
where $L_{u,v}^D$ and $L_{v,u}^D$ are the total losses on the edges in the path from $u$ to $v$ in $G_C$ downward of vertex $u$ and $v$ respectively. These losses can be fully written in binary variables and load arc variables, as shown in \cite{silva_qubo_2023}.

The loss term is then defined as
\begin{equation}
    \mathcal{L} = \sum_{(u,v)\in E_0} L_{u,v}^D+L_{v,u}^D.
\end{equation}

The total HUBO cost function is a weighted sum of feasibility-enforcing penalty terms and a physical loss term,
\begin{align}
\mathrm{Obj}_{\text{HUBO}} =\;&
\lambda_{\text{vertex}} \cdot C_{\text{vertex}} \Ctag \label{eq:c_vertex}\\
+&\lambda_{\text{edge}} \cdot C_{\text{edge}} \Ctag \label{eq:c_edge}\\
+&\lambda_{\text{cycle}} \cdot C_{\text{cycle}} \Ctag \label{eq:c_cycle}\\
+&\lambda_{\text{path}} \cdot C_{\text{path}} \Ctag \label{eq:c_path}\\
+&\lambda_{\text{implies}} \cdot C_{\text{implies}} \Ctag\label{eq:c_implies}\\
+&\lambda_{\text{loss}} \cdot \mathcal{L}, \Otag \label{eq:c_loss}
\end{align}
where in equations \eqref{eq:c_vertex} - \eqref{eq:c_implies}, the $C_{i}$'s denote the constraint cost terms introduced in the previous subsections, $\mathcal{L}$ in equation \eqref{eq:c_loss} represents the network power losses, and $\lambda_i$ are their corresponding (penalty) weights. Note that we don't choose values for $\lambda$'s in this work, as we only count the number of terms.

\subsection{Quantum Resource Estimation}
\subsubsection{HUBO to Cost Operator}
To map the HUBO formulation of the network reconfiguration problem to a quantum algorithm, we need to implement the HUBO as a cost operator $U_{HUBO}$. This, together with a mixer operator $U_{mix}$, can later be used as one optimisation layer $U_{opt}$ in quantum optimisation algorithms like the Quantum Approximate Optimisation Algorithm (QAOA) \cite{farhi_quantum_2014} or the bias-field Digitised Counterdiabatic Quantum Optimisation (bf-DCQO) algorithm \cite{romero_bias-field_2025}. Starting from the HUBO, we map each binary variable $x_i \in \{0,1\}$ to a spin variable $z_i \in \{-1, 1\}$ via $x_i = (1-z_i)/2$. The HUBO cost function is then converted into a Pauli Hamiltonian $\mathcal{H}_{\text{HUBO}}$, where each term corresponds to a tensor product of Pauli $Z$ operators acting on the qubits associated with the variables in that term, i.e.,
\begin{equation}
\mathcal{H}_{\text{HUBO}} = \sum_{i} c_i \bigotimes_{j \in S_i} Z_j,
\end{equation}
where $S_i$ is the set of qubits corresponding to variables in term $i$ and $c_i$ is its coefficient. Now the cost operator is $U_{HUBO}=e^{-i \gamma \mathcal{H}_{\text{cost}}}$, and since Pauli $Z$ operators commute, we can write
\begin{equation}
    U_{HUBO}=e^{-i \gamma \mathcal{H}_{\text{cost}}} =  e^{-i\gamma \sum_{i} c_i \bigotimes_{j \in S_i} Z_j} =  \prod_i e^{-i\gamma c_i \bigotimes_{j \in S_i} Z_j}.
\end{equation}
For resource estimation, we do not need to explicitly create the operator, as we can directly count the number of Pauli $Z$-rotations. Then we add two rotation gates per qubit to implement a mixer operator to get $U_{opt}$.
Because the number of times the optimisation layer is applied is very specific to the algorithm and dependent on the classical optimisation routine, we are more interested in the resources required to implement \textit{one} layer of a cost operator and mixer operator.

\begin{table}[h]
\centering
\caption{Circuit statistics for implementing one optimisation layer for the HUBO formulation of selected biconnected components from the Alliander and benchmarking dataset. Reported metrics include the number of nodes and edges, in $G_C$ and $G_0$, total HUBO interaction terms, required logical qubits, and logical rotation gates. Note that the 33-node IEEE test network has a 32-node biconnected component.}
\label{tab:hubo_qaoa_biconnected}
\small
\resizebox{\textwidth}{!}{%
\begin{tabular}{llllll}
\hline
\textbf{Component $G_C$} &
\textbf{\# Nodes $G_C$ / $G_0$} &
\textbf{\# Edges $G_C$ / $G_0$} &
\textbf{\# Interactions} &
\textbf{\# Logical Qubits} &
\textbf{\# Logical Rotation Gates} \\
\hline
IEEE \cite{ieee-33} & 32 / 9 & 36 / 13 & 33,616 & 667 & 59,296 \\
\hline
Arnhem-0  & 7 / 3 & 7 / 3 & 42 & 14 & 94 \\
Arnhem-1 & 13 / 3 & 13 / 3& 150 & 17 & 272 \\
Arnhem-2 & 14 / 4 & 15 / 5 & 607 & 53 & 1,110 \\
Arnhem-3 & 309 / 65 & 407 / 108 & 408,029,626 & 61,172 & 494,437,177 \\
\hline
\end{tabular}
}
\end{table}

\subsubsection{Resource Estimation Procedure}
For fault-tolerant quantum resource estimation, we follow \cite{beverland_assessing_2022} and use the Microsoft Quantum Resource Estimator to estimate the required number of physical qubits and runtimes for six typical hardware types: four gate-based and two Majorana qubits. The estimator translates logical circuit requirements into physical resource estimates by accounting for error correction with a surface code, where each logical qubit is encoded into many physical qubits to achieve fault-tolerant operation. The surface code distance is determined automatically by the given logical requirements and qubits, see \cite{beverland_assessing_2022} for details.

\section{Results}

Table~\ref{tab:hubo_qaoa_biconnected} summarises the structural properties of selected biconnected components from the Alliander dataset and the corresponding logical requirements derived from the HUBO formulation. The results show that even moderately sized grid components lead to a rapid growth in the total number of HUBO interaction terms. For the smallest biconnected component (Arnhem-0) in Figure \ref{fig:arnhem-MV}, the HUBO model contains 42 interaction terms, requiring 14 logical qubits and 94 logical rotation gates. However, as the number of nodes and edges increases, the number of interaction terms grows extremely. For example, Arnhem-2 (14 nodes, 15 edges) results in 607 interaction terms, requiring 53 logical qubits and 1, 110 logical rotation gates, while Arnhem-3 (309 nodes, 407 edges) results in over $4.08\cdot 10^8$ interaction terms, requiring 61,172 logical qubits and over $4.94\cdot 10^8$ logical rotation gates.

Comparison of the number of nodes and edges between biconnected components shows that components with more edges produce substantially more interaction terms. Since the number of edges provides a proxy for network connectivity and cyclicity, these results suggest that circuit size is driven not only by the number of nodes but also by the structural complexity of the grid through the combinatorial growth of higher-order HUBO interactions.

\begin{figure}[t]
\centering
\includegraphics[width=0.7\textwidth]{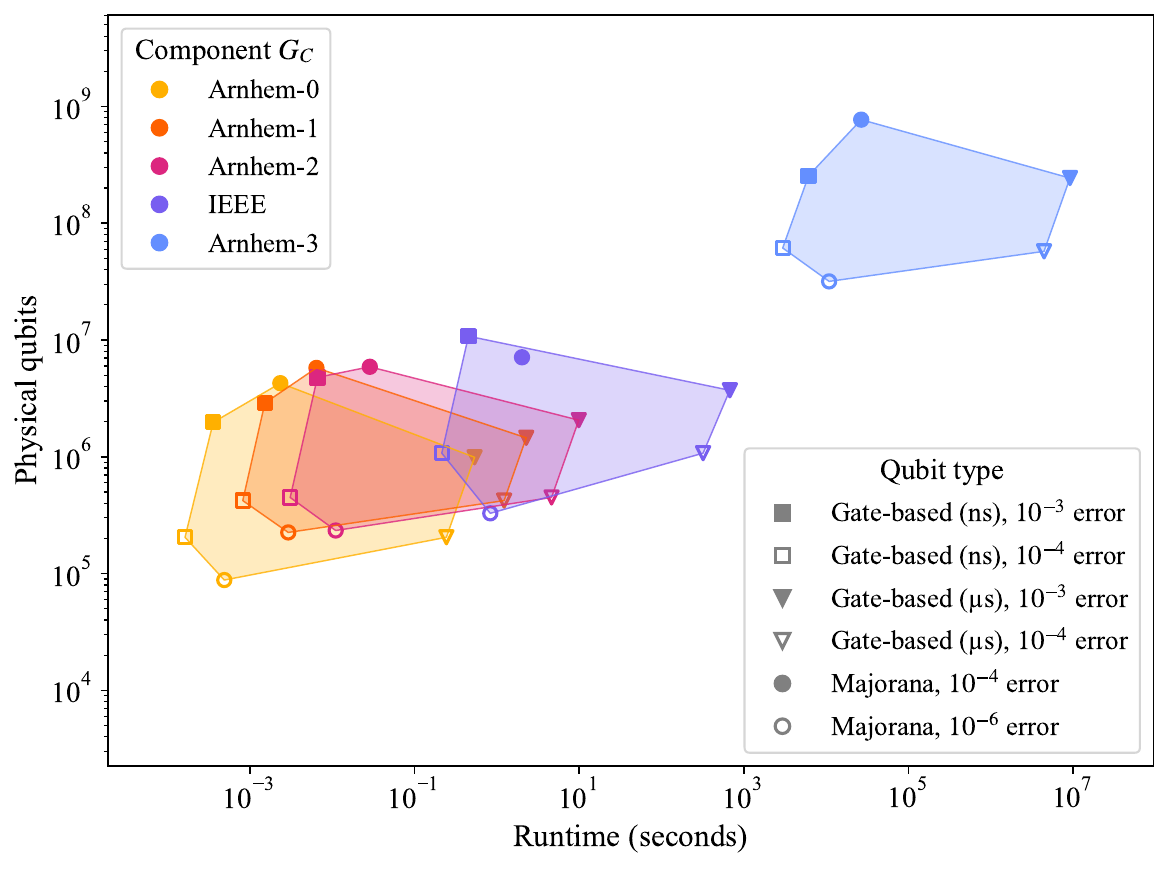}
\caption{Resource estimation results for implementing one optimisation layer $U_{opt}$. Each marker represents a qubit technology configuration. The shaded region highlights the achievable trade-off region between runtime and required physical qubits. Microsoft's Quantum Resource Estimator determines surface code distance automatically based on the logical requirements. It should be noted that these are not the resources required for a full optimisation pipeline, which would mean multiple applications of $U_{opt}$ and adding read-out time.}
\label{fig:resource_estimation}
\end{figure}

Figure~\ref{fig:resource_estimation} presents the physical resource estimates for implementing one optimisation layer under different qubit technology assumptions. Each marker represents a specific technology configuration, while the shaded region indicates the achievable trade-off between runtime and required physical qubits, which represent the fault-tolerant implementation of the logical circuits. It is clear that the runtime estimates depend on the number of interactions in the HUBO. Also, the actual worst-case runtime estimates are in the order of $10^7s$. Given that this reflects the runtime for applying 1 cost operator $U_{HUBO}$, the total runtime of a quantum algorithm could grow to order $10^8-10^9$ seconds, depending on the quantum algorithm. This shows the need for good quantum optimisation algorithms aided by good classical optimisers for optimising the variational circuit parameters.

\section{Discussion \& Further Research}
This study explores the feasibility of applying gate-based quantum computing to real-world distribution networks. Several limitations remain. Due to system size, the HUBO formulation was not explicitly solved and should be validated on smaller networks with known solutions in future work. Future work should also include the real electrical parameters of our grids, which were omitted in this research because they do not affect QRE but are essential for solving the optimisation problem.

In addition, the performed QRE should be used as an initial assessment. Practical optimisation requires repeated applications of the cost and mixer operators, and future work should evaluate the resources required for a complete optimisation pipeline. More refined estimates could also incorporate more refined techniques, such as advanced quantum error correction or NISQ-based approaches. Lastly, a direct comparison between HUBO and QUBO formulations would further clarify trade-offs in quantum resource requirements, such as qubit requirements and circuit depth.

Beyond methodological research extensions, future work could also consider the development and incorporation of Smart Grids within the proposed optimisation framework. In modern distribution networks, remotely controlled switching devices enable dynamic reconfiguration of network topologies. Such capabilities, which are already employed by Portugal for example, allow distribution systems to adapt in near real-time to changing operating conditions and have been shown to increase the integration potential of renewable energy sources through DNR \cite{lueken2012distribution}. Additionally, the incorporation of distributed energy sources such as wind and solar introduces additional dynamics and complexity, making DNR a frequently repeated or near real-time optimisation task. Therefore, extending the HUBO formulation to integrate multiple time-dependent sources of generation and demand, and quantum optimisation may play an important role in enabling such higher-complexity, larger-scale implementations.

\section{Conclusion}
Our work demonstrates the importance of evaluating quantum optimisation techniques on real-world distribution networks. Focussing on an MV network of Alliander and a benchmarking data set, the HUBO formulation was found for various biconnected components. Logical counts to implement one cost operator and one mixer operator are calculated. The results suggest that resource requirements may depend not only on the number of nodes and edges but also on graph characteristics such as connectivity and cyclicity, which are results of the increased number of nodes and edges. Together, these all increase the number of interaction terms, which subsequently influences the estimated runtimes. Despite the limitations of our work, it shows that it is feasible to construct HUBO formulations and quantum circuits for real-world MV networks. This paves the way forward to future work on quantum optimisation for DNR and power loss reduction in electricity grids.

\section*{Acknowledgements}
We would like to acknowledge the contributions from Niels Vercauteren, Nikki Jaspers, Julian van Velzen, Kerwin Buijsman, James Cruise, and Walden Killick. Their insights and support have been valuable to this study in various ways. This research was supported by Alliander DSO, with additional contributions from the Capgemini Quantum Lab. This work was performed using the ALICE compute resources provided by Leiden University.

\enlargethispage{2\baselineskip}
\bibliographystyle{ieeetr}
\bibliography{references}

\end{document}